# Coupled Spin and Pseudo-magnetic Field in Graphene Nanoribbons

Wen-Yu He, Lin He*

Department of Physics, Beijing Normal University, Beijing, 100875, People's Republic of China

**Pseudo-magnetic field becomes an experimental reality after the observation of zero-field Landau level-like quantization in strained graphene, but it is not expected that the time-reversal symmetric pseudo-magnetic fields will have any effect on the spin degree of freedom of the charge carriers. Here, we demonstrate that spin-orbit coupling (SOC) could act as a bridge between pseudo-magnetic field and spin. In quantum spin Hall (QSH) states, the direction of the spin of edge states is tied to their direction of motion because of the SOC. The pseudo-magnetic field affects the clockwise and counter-clock-wise edge currents of the QSH states, and consequently lifts the degenerate edge states of opposite spin orientation. Because of opposite signs of the pseudo-magnetic field in two valleys of graphene, the one-dimensional charge carriers at the two opposite edges have different group velocities, and in some special cases the edge states can only propagate at one edge of the nanoribbon and the group velocity at the other edge becomes zero.**

Generally speaking, we can divide the effects of a real magnetic field on charge carriers into two parts: the orbital field and the Zeeman field. The orbital field influences the electronic motion by introducing unimodular phase factors in the electron hopping amplitudes, and the Zeeman field lifts the energy degenerate states of opposite spin orientation. In graphene, a strain-induced hopping modulation between sublattices affects the Dirac fermions like an effective gauge field (pseudo-magnetic field) and can result in partially flat bands in the band structure of graphene at discrete energies, which are the analog of Landau levels in real magnetic fields [1-3]. Such pseudo-magnetic fields become an experimental reality after the observation of zero-field Landau level-like quantization in strained graphene [4-8]. The slight deference between the pseudo-magnetic field and the orbital magnetic field is that the pseudo-magnetic field preserves time reversal symmetry and has opposite signs for charge carriers in the two low-energy valleys, K and K′. Because of lacking the Zeeman term, therefore, it is not expected that the pseudo-magnetic field will have any effect on the spin degree of freedom of the charge carriers [2,3].

Here, we present theoretical investigations of the pseudo-magnetic fields in strained graphene, and we show that the pseudo-magnetic fields can affect the spin degree of freedom through spin-orbit coupling, which links spin and momentum of the charge carriers. Our result indicates that the pseudo-magnetic fields can lift the degenerate edge states of opposite spin orientation in zigzag graphene nanoribbons and result in asymmetry of the quantum spin Hall (QSH) states: the spin-polarized edge states have different group velocities at the two opposite edges, and in some special cases the edge states propagate without dissipation at one edge of the nanoribbon and the group velocity at the opposite edge becomes zero. Our result opens a new door to explore spin-based electronics in graphene.

The proposal of the QSH states (or the topological edge states) in graphene was one of the milestone works in the development of the field of topological insulator [9-14]. However, graphene's extremely weak intrinsic spin-orbit coupling makes the realization of the QSH states to be practically unrealistic [15,16]. Recently, recipes for enhancing spin-orbit interaction on graphene by introduction of adatoms have been suggested [17-20]. Experimentally, it was demonstrated that small amounts of covalently bonded hydrogen atoms induce a colossal enhancement of the spin-orbit coupling in graphene by three orders of magnitude, i.e., the spin-orbit coupling increases from the order of $10^{-3}$ meV to about 2.5 meV [21]. Thus, the experimental study of graphene with a moderate strength of spin-orbit interaction now appears feasible, and it becomes possible to realize the QSH states in graphene at an experimentally accessible temperature.

In the tight binding model of graphene with the spin-orbit interaction, the Hamiltonian takes the form [9,10],

$$H = \sum_{\langle ij\rangle\alpha} t c^{\dagger}_{i\alpha} c_{j\alpha} + \sum_{\langle\langle ij\rangle\rangle\alpha\beta} it_2 v_{ij} s^z_{\alpha\beta} c^{\dagger}_{i\alpha} c_{j\beta}. \qquad (1)$$

Here, the first term is the nearest neighbor hopping term on the honeycomb lattice, $t$ is the hopping integral, and the operators $c^{\dagger}_{i\alpha}$ ($c_{i\alpha}$) create (annihilate) an electron with spin $\alpha$ at site $i$. Around the Dirac points, the first term becomes two-dimensional massless Dirac equation, which describes the low-energy behaviors of the charge carriers in graphene monolayer [1]. The second term of Hamiltonian (1) describes the spin-orbit interaction by introducing a spin dependent second neighbor hopping $t_2$. Here, $v_{ij} = \pm 1$ depending on the orientation of the two nearest neighbor bonds the electron traverses in going from site $j$ to $i$, and $s_{\alpha\beta}^z$ is the Pauli matrix describing the electron's spin [9,10]. The QSH states can be obtained by solving Hamiltonian (1) in a nanoribbon geometry (see Figure S1 in supporting materials [22]). In the QSH states, the electrons with opposite spin orientation propagate in opposite directions at each edge, and the electrons with the same spin orientation between the two opposite edges also propagate in opposite directions [9,10]. For a zigzag graphene nanoribbon, the QSH states become a non-topological surface states when the spin-orbit coupling is vanishingly small [9,23].

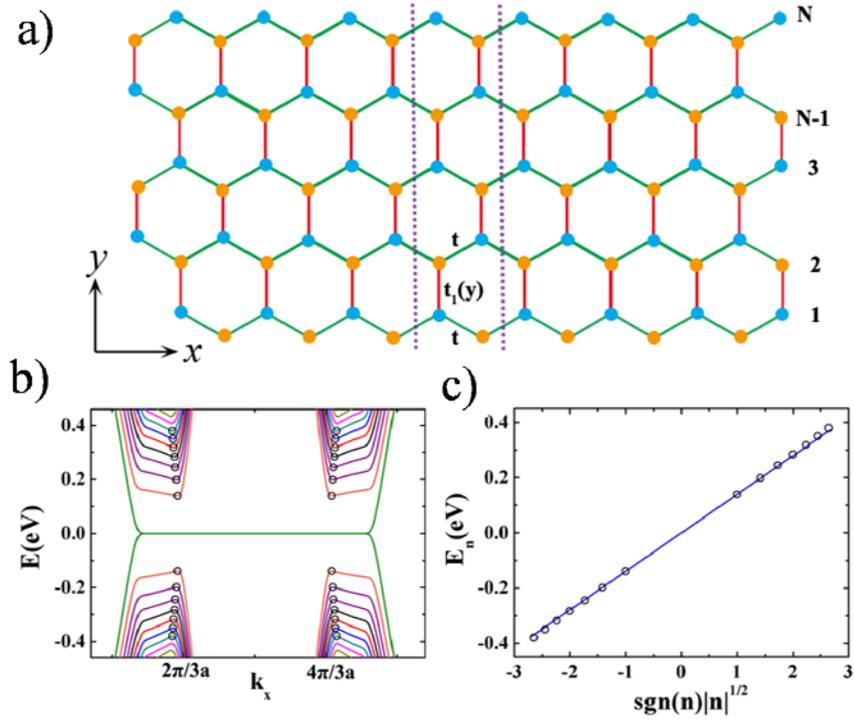

**Figure 1**. A Zigzag graphene nanoribbon in uniform pseudo-magnetic fields. (**a**). Diagram of a zigzag graphene nanoribbon. The hopping matrix element along y direction (red bonds) is changed to induce the gauge field $A_x = B_S y$ and $A_y = 0$. N is the number of zigzag chains in the nanoribbon. (**b**). Energy dispersion for the graphene nanoribbon in (**a**). The strain-induced Landau-level-like flat bands acquire a small linear dispersion because of the hybridization with the non-topological surface states of the zigzag nanoribbon. (**c**). A comparison between energies of the black open circles in (**b**) and the Landau levels generated in real magnetic fields. The blue line is calculated according to $E_n = \text{sgn}(n)\hbar\omega_c\sqrt{|n|}$, where $n$ is the index of the Landau levels and $\omega_c = \sqrt{2e\hbar v_F^2 B_S}$ with $v_F$ as the Fermi velocity and the pseudo-magnetic fields $|B_S| = 15.2$ T.

For the case that only the first term of Hamiltonian (1) is taken into account, the strain-induced lattice deformations change the electron hopping between sublattices and give rise to an effective gauge field **A** in the Dirac equation. Certain spatially varying deformation patterns affect the Dirac fermions in graphene mimicking the effect of uniform pseudo-magnetic fields $B_S = \nabla \times \mathbf{A}$. Figure 1 shows a concrete lattice to realize a uniform pseudo-magnetic field with $A_x = \frac{c}{ev_F}(t - t_1) = B_S y$ and $A_y = 0$. The pseudo-magnetic field cannot lift the degenerate band of opposite spin orientation, and we obtain a twofold degenerate band structure even when the spin degree of freedom is considered. For the zigzag graphene nanoribbon, the pseudo-magnetic field have opposite signs in the two valleys (K and K′), which plays a vital role in the emergence of the asymmetric QSH states.

When both the pseudo-magnetic fields and the spin-orbit coupling are taken into account, the pseudo-magnetic fields lift the degenerate edge states of opposite spin orientation through the spin-orbit coupling, which links the spin and momentum of the charge carriers. Consequently, we obtain a new and unique QSH state in zigzag graphene nanoribbons. Figure 2 shows the one-dimensional energy bands for a zigzag graphene nanoribbon. The electronic states of opposite spin orientation are still degenerate in all "bulk" bands and the valley-dependent pseudo-magnetic fields only lift the degenerate edge states. The energy-level shifts are of opposite signs for the degenerate edge states in the two valleys. For the pseudo-magnetic fields used in the calculation, the energies of edge states with spin up (spin down) are raised (lowered) for the states with $k_x < \pi/a$ (here, $\pi/a$ is the Brillouin zone boundary), i.e., in the K valley. The energies of edge states with spin up (spin down) are lowered (raised) for the states with $k_x > \pi/a$, i.e., in the K′ valley. If we switch the direction of the pseudo-magnetic fields in the K valley, then the energies of edge states with spin up (spin down) are lowered (raised) for the states with $k_x < \pi/a$.

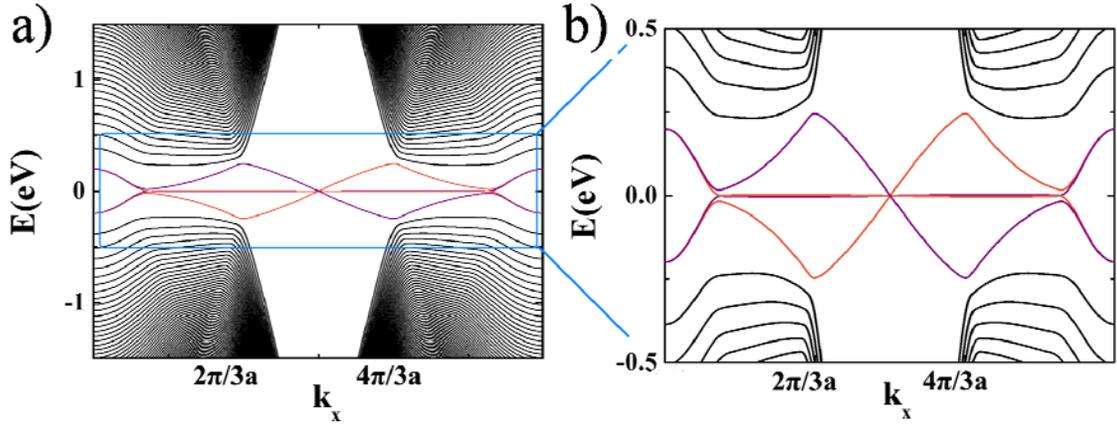

**Figure 2.** One-dimensional energy dispersion for a zigzag graphene nanoribbon with N = 600. (**a and b**) Both the uniform pseudo-magnetic fields $|B_S|$ = 25.90 T and the spin-orbit coupling ($t_2/t$ = 0.05) are taking into account in the calculation. Here, $B_S$ = 25.90 T in the K valley and $B_S$ = - 25.90 T in the K′ valley (we define direction of the pseudo-magnetic field "down" as positive; "up", negative pseudo-magnetic field). The orange color denotes the spin down edge states; violet, spin up edge states. Panel (**b**) shows the low-energy band in panel **a**. Because of opposite signs of the pseudo-magnetic fields in the two valleys, the pseudo-magnetic fields induce energy-level shifts of opposite signs for the degenerate edge states of opposite spin orientation. One spin up edge state and one spin down edge state become perfectly flat.

To understand the physics behind this phenomenon, we have carried out the same calculation in a static orbital magnetic field (the Zeeman field is not included in the calculation and the direction of the magnetic field is the same in the two valleys). With introducing the orbital magnetic field, the hopping integral $t_{ij}$ is replaced by $t_{ij}e^{i2\pi\varphi_{i,j}}$, where $\phi_{i,j}$ is given by the line integral of the vector potential from site $i$ to $j$. In both valleys, the energy of the clockwise edge states (spin up electrons) is raised and the energy of the counter-clock-wise edge states (spin down electrons) is lowered by the orbital magnetic field, as shown in Figure 3. This result indicates that the spin-split edge states are induced by the interplay between the orbital magnetic field (or the pseudo-magnetic field) and the clockwise and counter-clock-wise edge currents (Figure 3). The spin-orbit coupling links the spin and the momentum, the spin of an electron that has been affected is determined by its momentum (whether the edge currents is clockwise or counter-clock-wise). Therefore, the spin-orbit coupling acts as a bridge between the pseudo-magnetic field and the spin degree of freedom. For armchair graphene nanoribbons, the effect of pseudo-magnetic fields on the edge states is perfectly canceled in the K and K′ valleys (see Figure S2 of supporting materials [22]). Therefore, the asymmetric QSH effect can only observed in zigzag graphene nanoribbons.

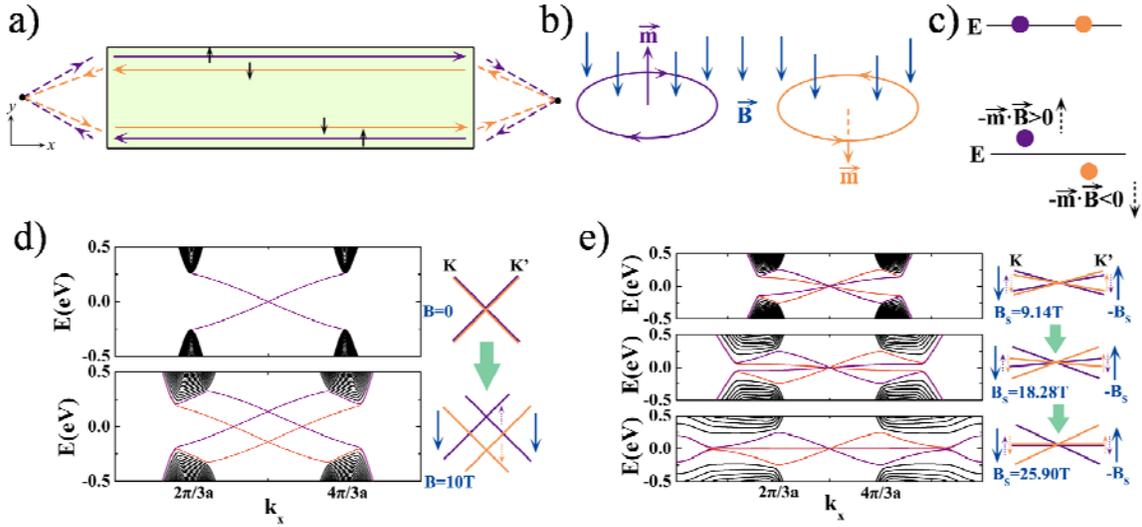

**Figure 3.** Origin of the asymmetric QSH states in graphene nanoribbons. **a).** Schematic diagram showing the QSH states in a graphene nanoribbon. The spin up and spin down electrons prograte clockwise and counter-clock-wise respectively. (**b** and **c**) The interplay between the orbital magnetic field (or the pseudomagnetic field) and effective magnetic moments generated by the clockwise and counter-clock-wise edge currents lifts the degeneracy of the edge states. **d).** Upper panel: energy dispersion of a zigzag graphene nanoribbon with N = 600 in zero magnetic field. Lower panel: energy dispersion of the zigzag graphene nanoribbon in a static orbital magnetic field B = 10 T. In both valleys, the energy of the clockwise edge states (spin up electrons) is raised and the energy of the counter-clock-wise edge states (spin down electrons) is lowered. **e).** The low-energy bands of the zigzag graphene nanoribbon in different values of the pseudo-magnetic fields.

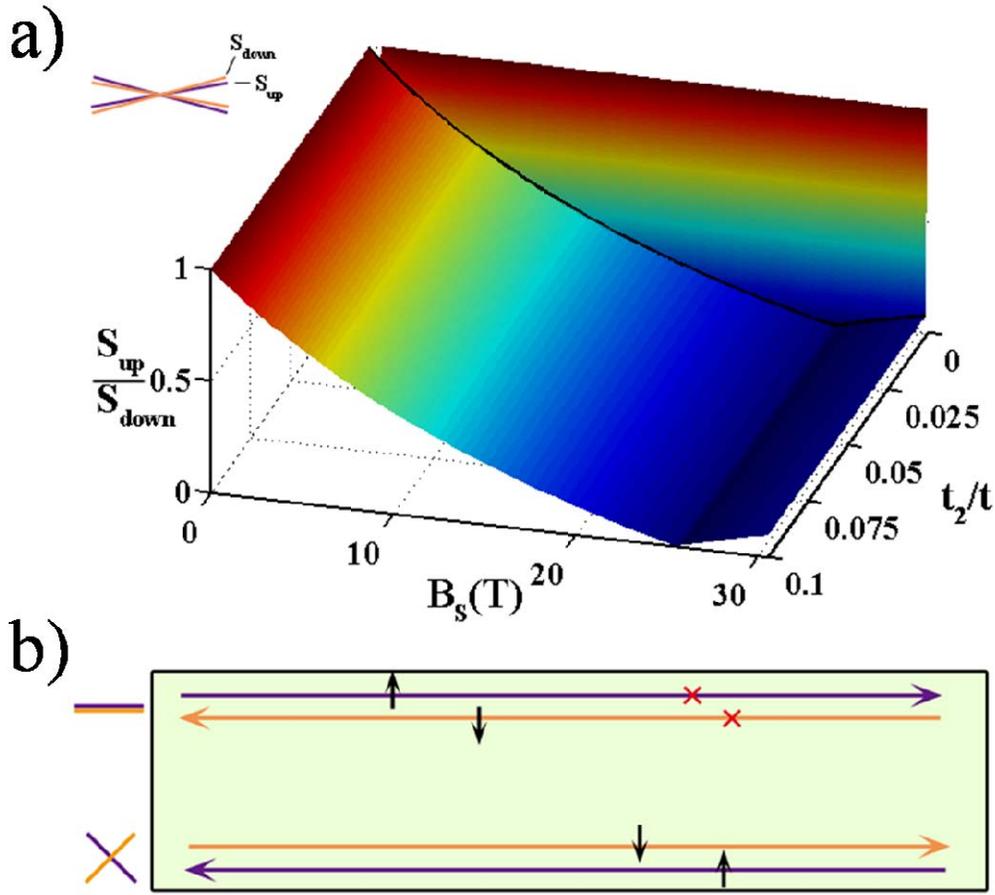

**Figure 4.** The asymmetic QSH states in a graphene nanoribbon. **a).** The ratio of $S_{up}/S_{down}$ as a function of the pseudo-magnetic fields and the strength of spin-orbit coupling. $S_{up}$ and $S_{down}$ are the slope of edge states in the K′ valley with spin up and spin down respectively. The solid black line is plotted to show the ratio of $S_{up}/S_{down}$ as a function of the pseudo-magnetic fields for $t_2/t = 0.001$. The system becomes asymmetric QSH state when $S_{up}/S_{down} = 0$. **b).** Schematic diagram showing the asymmetric QSH states in a graphene nanoribbon. The group velocity of the QSH states becomes zero at one edge of the nanoribbon.

The opposite signs of pseudo-magnetic fields in the K and K′ valleys of zigzag graphene nanoribbons lead to opposite signs of energy-level shifts for the degenerate edge states in the two valleys, as shown in Figure 3. This is the key factor for the emergence of the asymmetric QSH states. The pseudo-magnetic fields lead to asymmetric group velocities of the edge states at opposite edges of the nanoribbon. In some peculiar values of the pseudo-magnetic fields, both the spin up and spin down edge states at one edge of the nanoribbon become perfectly flat, as shown in Figure 2 and Figure 3. Then, the group velocity of the edge states becomes zero. It indicates that the QSH states only propagate dissipationless at one edge of the nanoribbon and the edge state is suppressed at the opposite edge, as shown in Figure 4. In QSH states, the edge states are not chiral because of that each edge has electrons which propagate in both directions and the counterpropagation of the same spin states at the two opposite edges. When the group velocity at one edge of the nanoribbon becomes zero, the graphene nanoribbon provides dissipationless spin-polarized current only at one edge, and the edge states become chiral. Then, the graphene nanoribbon acts as a dissipationless "spin battery" [24] and the edge states could be used as a dissipationless spin-filtering path for spintronic devices. The asymmetric QSH state distincts from other existing quantum Hall related effects (i.e., the quantum Hall effect, the QSH effect, and the quantum anomalous Hall effect) and adds a new member in the quantum Hall family [25].

In the preceding calculation, we used a uniform pseudo-magnetic field to induce the asymmetric QSH states in graphene nanoribbons. Actually, this is not necessary for the emergence of the asymmetric QSH states. The asymmetric QSH states also exist in zigzag graphene nanoribbons with non-uniform pseudo-magnetic fields (see Figure S3 of supporting materials [22]), and it could be observed in graphene nanoribbons with a small part of deformed structure (see Figure S4 of supporting materials [22]). The asymmetry of the edge states at opposite edges of the nanoribbon is robust when the net pseudo-magnetic field of the nanoribbon is not zero in a valley. The asymmetric QSH states can exist and the asymmetry is obvious for $t_2/t = 0.001$, as shown in Figure 4. It indicates that our theoretical results hold well for the strength of spin-orbit interaction in the range accessible by experiments [21]. By taking advantage of the development in the fabrication of graphene nanoribbons [26], we believe that the experimental detection of the asymmetric QSH state is possible. The asymmetric QSH states could be observable by studying low temperature charge transport [12,27] in weakly hydrogenated zigzag graphene nanoribbons with partially deformed structure. Other graphene sheet analogues with strong spin-orbit coupling [28,29] and lattice deformations should also exhibit this effect.

In summary, we show that the time-reversal symmetric pseudomagnetic fields can affect the spin of charge carriers through the spin-orbit coupling, and the pseudo-magnetic fields lift the degenerate edge states of opposite spin orientation in zigzag graphene nanoribbons. This effect ont only results in asymmetric QSH states in graphene nanoribbons but also opens a new door to explore spin-based electronics in graphene.

*Email: helin@bnu.edu.cn.

**Acknowledgements**

The authors would like to thank Hua Jiang, Cheng-cheng Liu, and Yugui Yao for helpful discussions. We are grateful to National Science Foundation (Grant No. 11004010), National Key Basic Research Program of China (Grant No. 2013CBA01603), and the Fundamental Research Funds for the Central Universities.